\documentclass[%
aps,
reprint, 
superscriptaddress,
amsmath,amssymb,
prc,
floatfix,
]{revtex4-1}

\usepackage{graphicx}
\usepackage{dcolumn}
\usepackage{bm}

\begin{document}

\title{Neutron emission following nuclear muon capture on palladium isotopes}

\author{T.~Y.~Saito}
 \email{takeshi.saito@riken.jp}
 \affiliation{Graduate School of Science, the University of Tokyo, 7-3-1 Hongo, Bunkyo-ku, 113-0033 Tokyo, Japan}
 \affiliation{Pioneering Research Institute, RIKEN, 2-1 Hirosawa, Wako-shi, 351-0198 Saitama, Japan}
\author{M.~Niikura}
 \affiliation{Graduate School of Science, the University of Tokyo, 7-3-1 Hongo, Bunkyo-ku, 113-0033 Tokyo, Japan}
 \affiliation{Nishina Center, RIKEN, 2-1 Hirosawa, Wako-shi, 351-0198 Saitama, Japan}
\author{T.~Matsuzaki}
 \affiliation{Nishina Center, RIKEN, 2-1 Hirosawa, Wako-shi, 351-0198 Saitama, Japan}
\author{S.~Abe}
 \affiliation{Japan Atomic Energy Agency (JAEA), 2-4 Shirakata, Tokai-mura, Naka-gun, 319-1195 Ibaraki, Japan}
\author{K.~Ishida} 
 \affiliation{Nishina Center, RIKEN, 2-1 Hirosawa, Wako-shi, 351-0198 Saitama, Japan}
\author{S.~Kawase}
 \affiliation{Interdisciplinary Graduate School of Engineering Sciences,
Kyushu University, Kasuga, Fukuoka 816-0811, Japan}
\author{Y.~Kawashima}
 \affiliation{Research Center for Nuclear Physics, Osaka University, 10-1 Mihogaoka, Ibaraki-shi, 567-0047 Osaka, Japan}
\author{T.~Koiwai}
 \affiliation{Graduate School of Science, the University of Tokyo, 7-3-1 Hongo, Bunkyo-ku, 113-0033 Tokyo, Japan}
\author{K.~Matsui}
 \affiliation{Graduate School of Science, the University of Tokyo, 7-3-1 Hongo, Bunkyo-ku, 113-0033 Tokyo, Japan}
\author{S.~Momiyama}
 \affiliation{Graduate School of Science, the University of Tokyo, 7-3-1 Hongo, Bunkyo-ku, 113-0033 Tokyo, Japan}
 \author{A. Nambu}
 \affiliation{Graduate School of Science, Osaka University, 1-1 Machikaneyama-cho, Toyonaka-shi, 560-0043 Osaka, Japan}
\author{H.~Otsu}
 \affiliation{Nishina Center, RIKEN, 2-1 Hirosawa, Wako-shi, 351-0198 Saitama, Japan}
\author{H.~Sakurai}
 \affiliation{Nishina Center, RIKEN, 2-1 Hirosawa, Wako-shi, 351-0198 Saitama, Japan}
 \affiliation{Graduate School of Science, the University of Tokyo, 7-3-1 Hongo, Bunkyo-ku, 113-0033 Tokyo, Japan}
\author{A.~Sato}
 \affiliation{Graduate School of Science, Osaka University, 1-1 Machikaneyama-cho, Toyonaka-shi, 560-0043 Osaka, Japan}
 \affiliation{Research Center for Nuclear Physics, Osaka University, 10-1 Mihogaoka, Ibaraki-shi, 567-0047 Osaka, Japan}
\author{X.~Sun}
 \affiliation{Nishina Center, RIKEN, 2-1 Hirosawa, Wako-shi, 351-0198 Saitama, Japan}
 \author{A.~Taniguchi}
\affiliation{Institute for Integrated Radiation and Nuclear Science, Kyoto University, 2 Asashiro-Nishi, Kumatori-cho, Sennan-gun, 590-0494 Osaka, Japan}
\author{D.~Tomono}
 \affiliation{Research Center for Nuclear Physics, Osaka University, 10-1 Mihogaoka, Ibaraki-shi, 567-0047 Osaka, Japan}
\author{H.~Wang}
 \affiliation{Nishina Center, RIKEN, 2-1 Hirosawa, Wako-shi, 351-0198 Saitama, Japan}
\author{Y.~Watanabe}
 \affiliation{Nishina Center, RIKEN, 2-1 Hirosawa, Wako-shi, 351-0198 Saitama, Japan}
\author{K.~Wimmer}
 \affiliation{Graduate School of Science, the University of Tokyo, 7-3-1 Hongo, Bunkyo-ku, 113-0033 Tokyo, Japan}

\date{\today}

\begin{abstract}

\begin{description}
\item[Background]
Neutron emission is the dominant deexcitation process of the compound nucleus following nuclear muon capture.
However, systematic experimental data on neutron energy spectra and neutron-neutron angular correlations are sparse, limiting the understanding of reaction dynamics.
\item[Purpose]
The energy spectra of the neutrons emitted following nuclear muon capture on palladium isotopes ($A=104$, 105, 106, 108, and 110) were measured using isotopically enriched target.
\item[Method]   
The experiment was performed at the MuSIC-M1 beamline at the Research Center for Nuclear Physics (RCNP), Osaka University.
The neutrons and $\gamma$ rays were detected with twenty-one liquid scintillators and BaF$_2$ detectors. 
The time-of-flight method was used to determine the neutron energy.
\item[Results]
Neutron energy spectra from 1\,MeV up to 20\,MeV were measured for five palladium isotopes, providing the first systematic data in the $A\sim100$ region. 
The spectral shapes were compared with the previous measurement for heavy nuclei and theoretical calculations.
The neutron-neutron opening angle distribution was also measured and an indication of small angle correlation was found.
\item[Conclusions]
The spectral shape below 4\,MeV was well  explained consistently with the previous measurement by the evaporation model introducing a mass number scaling.
The neutron energy spectrum around 10\,MeV plays a key role in understanding the dynamics of the nuclear muon capture reaction because it is the result of the transition from the direct and pre-equilibrium neutron emission onto the evaporation process.    
\end{description}
\end{abstract}

\maketitle


\section{Introduction}\label{sec_int}

A negative muon stopped in matter forms an atomic bound state with a nucleus, known as a muonic atom.
The atomic ground state of the muonic atom predominantly decays via two weak processes: $\mu$-e decay and nuclear muon capture.
The nuclear muon capture is the conversion process of a proton in the nucleus and a negative muon in the atomic orbit to a neutron and a muon neutrino.
While this process is similar to electron capture, the large mass of the capturing lepton, the muon (105.6583755\,MeV/c$^2$\cite{Navas2024-prd}), results in a large reaction $Q$ value of approximately 100\,MeV.
Consequently, this large $Q$ value leads to the formation of an excited daughter nucleus, $(Z-1, A)^{*}$, following capture on a $(Z, A)$ nucleus, where $Z$ and $A$ denote the atomic and mass numbers, respectively.
The daughter compound nucleus predominantly deexcites by emitting particles such as neutrons and $\gamma$ rays, and sometimes light charged particles, such as protons, deuterons, and alpha particles.
In medium-heavy nuclei, charged particle emission is hindered by the Coulomb barrier, resulting in neutrons being the primary emitted particles.
Our recent experimental study on nuclear muon capture for palladium (Pd) isotopes~\cite{Niikura2024-bz} has quantitatively revealed that the majority of the reaction products (totaling approximately 90\% for muon capture of $^{106, 108, 110}$Pd) are from neutron emission.
In contrast, the production branching ratios for charged particle emission are very small, remaining below 1\%.
Therefore, measurements of the energy spectra of the emitted neutrons are crucial to understanding the dynamics of the excitation and decay processes on the nuclei following nuclear muon capture. 

Nuclear muon capture proceeds through three primary processes: direct, pre-equilibrium, and evaporation, and neutrons emitted from each of these processes exhibit different energy spectra.
Thus, the neutron energy spectra provide vital information for dissecting and understanding the contributions of these processes.
Furthermore, if multiple neutrons are emitted in the reaction, measurements of their angular correlation could provide a deeper understanding of the microscopic structure of the excited states.
However, experimental data on neutron emission following nuclear muon capture are relatively sparse, particularly in the medium-heavy mass region.
Previous studies on neutron energy spectra have focused mainly on light nuclei such as carbon, oxygen~\cite{Plett1971-dm, Eramzhyan1973-iz, Van_Der_Schaaf1983-an}, sulfur, and calcium~\cite{Evseev1970-js}, or heavy nuclei such as thallium, lead, and bismuth~\cite{Schroder1974-jb}.
These studies often used targets with their natural isotope abundance, which limits the discussion of isotope dependence.
The lack of systematic data in the medium-heavy region, along with limited studies on neutron angular correlations, highlights the need for further experimental investigation. 

The present study aims to systematically measure the energy spectra and angular correlations of neutrons emitted from muon capture of isotopically enriched palladium targets ($A=104$, 105, 106, 108, and 110).


\section{Experiment}\label{sec_exp}

The experiment was performed at the MuSIC-M1 beamline at the Research Center for Nuclear Physics (RCNP), Osaka University~\cite{Cook2013-sw, Matsumoto2015-mp, Cook2017-bd}
A graphite pion production target was irradiated with a 392-MeV primary proton beam accelerated by two cyclotrons.
The primary beam intensity was 1.1\,$\mu$A.
Negative pions produced at the production target were collected and transported by the pion capture solenoid, where they decayed into negative muons.
These muons were momentum-selected by two dipole magnets in the MuSIC-M1 beamline.
The muon momentum of 50\,MeV/c was selected and decelerated by a 5-mm-thick carbon degrader placed between upstream beamline detectors, which will be explained below. 
The combination of the beam momentum and the degrader thickness was optimized to maximize the number of muons stopped at the target.
The beam had a 50\,mm diameter in the full-width at half-maximum (FWHM), and its momentum spread was approximately 8.8\%.
The typical muon beam rate at the target was approximately 100 particles per second.
A Wien filter was used to remove contaminating electrons from the beam.

Five isotopically-enriched palladium targets ($A=104$, 105, 106, 108, and 110) were irradiated with the muon beam.
Each target had an approximate isotopic enrichment of 98\%.
These targets were also used in our previous experiments; further details, including their chemical and isotope compositions, material form, and weight, can be found in Table~I of Ref.~\cite{Saito2025-ds}.

\begin{figure}
  \centering
  \includegraphics[width = 8.6cm]
      {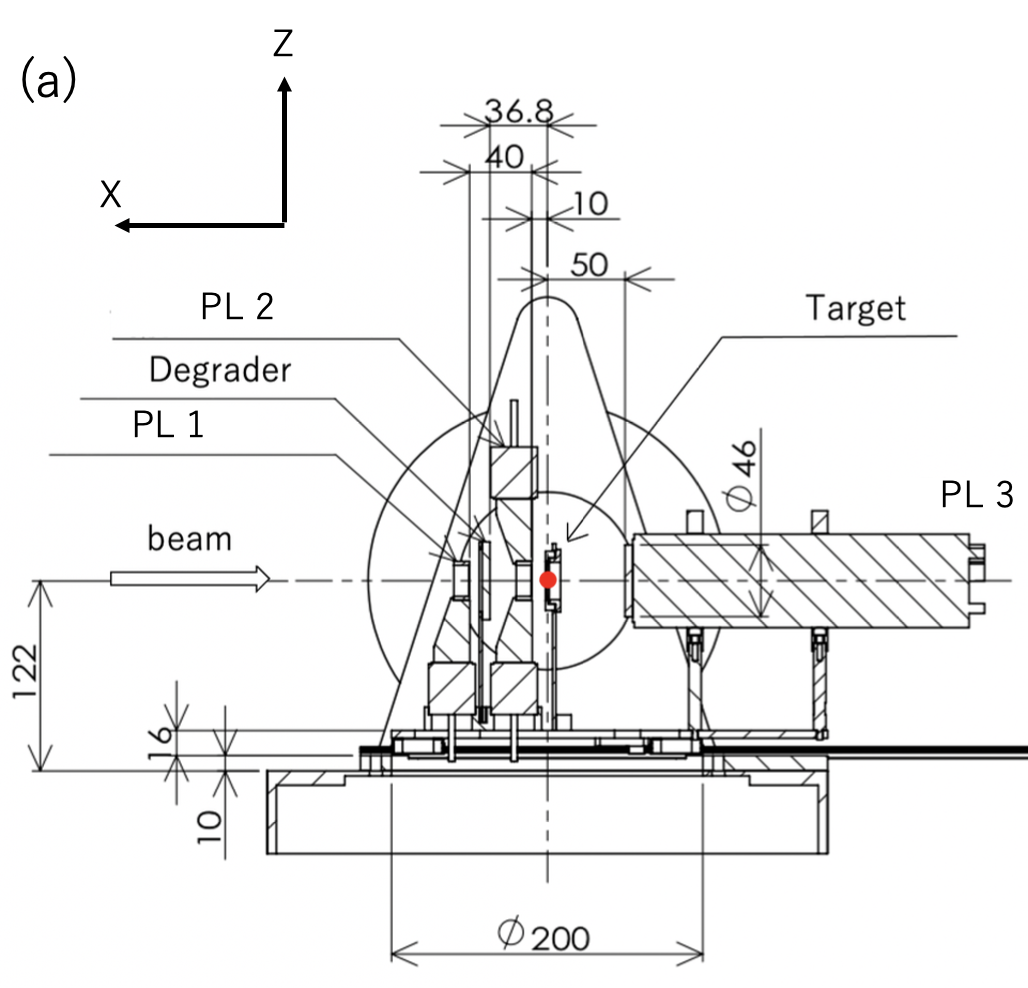}
  \includegraphics[width = 8.6cm]
      {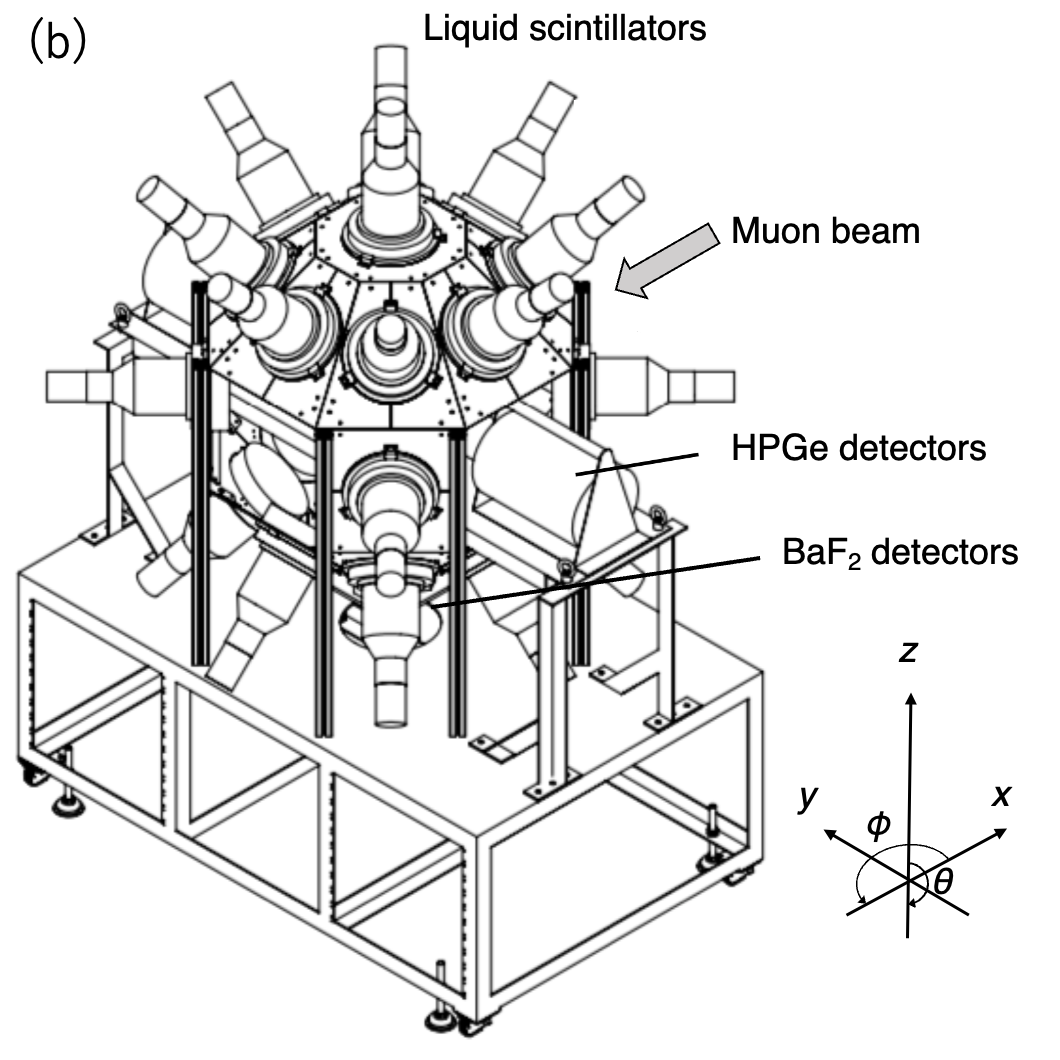}
  \caption{Technical drawing of the detectors.
  (a) Beamline detectors, target and degrader. Red dot represents the origin of the coordinate.
  (b) Detector array consists of twenty-one liquid scintillators, seven BaF$_2$ detectors, and two high-purity germanium (HPGe) detectors.}
  \label{fig:ex_detectors}
\end{figure}

Three plastic scintillation counters, referred to hereafter as PL1, PL2, and PL3, were used as beamline detectors. 
PL1 and PL2 were placed 5-cm and 1-cm upstream of the target, respectively, while PL3 was placed 5-cm downstream as shown in Fig.~\ref{fig:ex_detectors}(a).
The carbon degrader was located between PL1 and PL2, 3.7-cm upstream of the target.
PL1 and PL2 were square-shaped scintillators with dimensions of 20\,mm$\times$20\,mm and a thickness of 0.5\,mm, and PL3 was a disc-shaped scintillator with a diameter of 46\,mm and a thickness of 5\,mm.
Acrylic collimator was attached onto the upstream side of PL1 and PL2, with a hole size of 20\,mm in diameter, with a thickness of 10\,mm.
The scintillation photons were read by the photomultiplier tubes (PMTs: Hamamatsu H11934-100 for PL1 and PL2, and Hamamatsu H7195 for PL3).
PL1 had a single-sided readout from the bottom, PL2 had a double-sided readout from the top and bottom, and PL3 had a readout from the downstream side.

Figure~\ref{fig:ex_detectors}(b) shows the detector array used in the present neutron measurement following nuclear muon capture.
The array consisted of twenty-one liquid scintillators, seven BaF$_2$ detectors, and two high-purity germanium (HPGe) detectors.
The detector arrangement was as follows, according to the coordinate system shown in Fig.~\ref{fig:ex_detectors}.
In this system, the z-axis is vertically oriented upwards, the x-axis points upstream along the beam axis, $\theta$ denotes the polar angle, and $\phi$ represents the azimuthal angle.
The target was positioned at the origin of this coordinate system.
One liquid scintillator was placed at the top of the array ($\theta = 0^{\circ}$);
eight detectors were at $\theta = 45^{\circ}$ and $\phi = n\times45^{\circ}\,(n=0,1,2,...,7)$;
four detectors were horizontally installed at $\theta = 90^{\circ}$ and $\phi = n\times90^{\circ}+45^{\circ}\,(n=0,1,2,3)$;
and eight detectors were installed at $\theta = 135^{\circ}$ and $\phi = n\times45^{\circ}\,(n=0,1,2,...,7)$.
All liquid scintillators faced the target, with a distance of 40\,cm between the detector's surface and the target center.
The seven BaF$_2$ scintillators were placed at the bottom of the array ($\theta = 180^{\circ}$).
The two HPGe detectors were installed at $\theta = 90^{\circ}$ and $\phi=\pm 90^{\circ}$.

The liquid scintillators were used for neutron detection.
Each detector assembly included the organic liquid scintillator (Bicron BC-501A) and PMT (Hamamatsu H4144-01).
The BC-501A scintillator had pulse shape discrimination capability of neutrons and $\gamma$ rays.
Its effective volume was a cylindrical shape with a 200-mm diameter and a 50-mm thickness.
The dynamic range of the detectors was adjusted up to 10\,MeV$_{ee}$ (electron equivalent) using the $\gamma$ rays from the $^{60}$Co standard source.

Neutron energy from nuclear muon capture was measured using the time-of-flight (TOF) method.
Due to the finite lifetime of the muonic atom (approximately 100\,ns for muonic palladium), a time difference exists from the beam arrival to the nuclear muon capture reaction and subsequent neutron emission.
Consequently, an independent measurement of the start timing for the TOF measurement is required.
Seven BaF$_2$ detectors were used to define the start timing by detecting promptly emitted $\gamma$ rays from nuclear muon capture.
Each BaF$_2$ crystal had a hexagonal columnar shape narrowing toward the tip.
The height of the crystal was 10\,cm, and one edge of its hexagonal base measured 3.5\,cm. 
The dynamic range of the detectors was adjusted up to 10\,MeV photons.

The energy of the $\gamma$ and the muonic X rays was measured using the two HPGe detectors.
However, the HPGe detectors were not used in the present analysis.

The signals of the detectors were recorded by waveform digitizers (CAEN V1730), which operated with the WaveDump firmware.
This digitizer features 14-bit resolution and a 500\,MHz sampling frequency, corresponding to a 2\,ns sampling duration.
The record length was set to 300 samples (600\,ns) for the PL1, PL2, PL3, and BaF$_2$ detectors, and 1000 samples, (2\,$\mu$s) for the liquid scintillators.
The data acquisition was triggered by the coincidence of PL1 and PL2.
All the waveforms that coincided with the trigger were recorded; in other words, the data suppression algorithm against the low pulse heights was not utilized.

We measured 14.8-, 12.2-, 11.3-, 18.6-, and 13.8-hours data for $^{104,105,106,108,110}$Pd, respectively.
A reference measurement without a target (empty measurement) was also performed for 14.8 hours to estimate the background.

\section{Analysis}\label{sec_ana}

\begin{figure}
  \centering
  \includegraphics[width = 8.6cm]{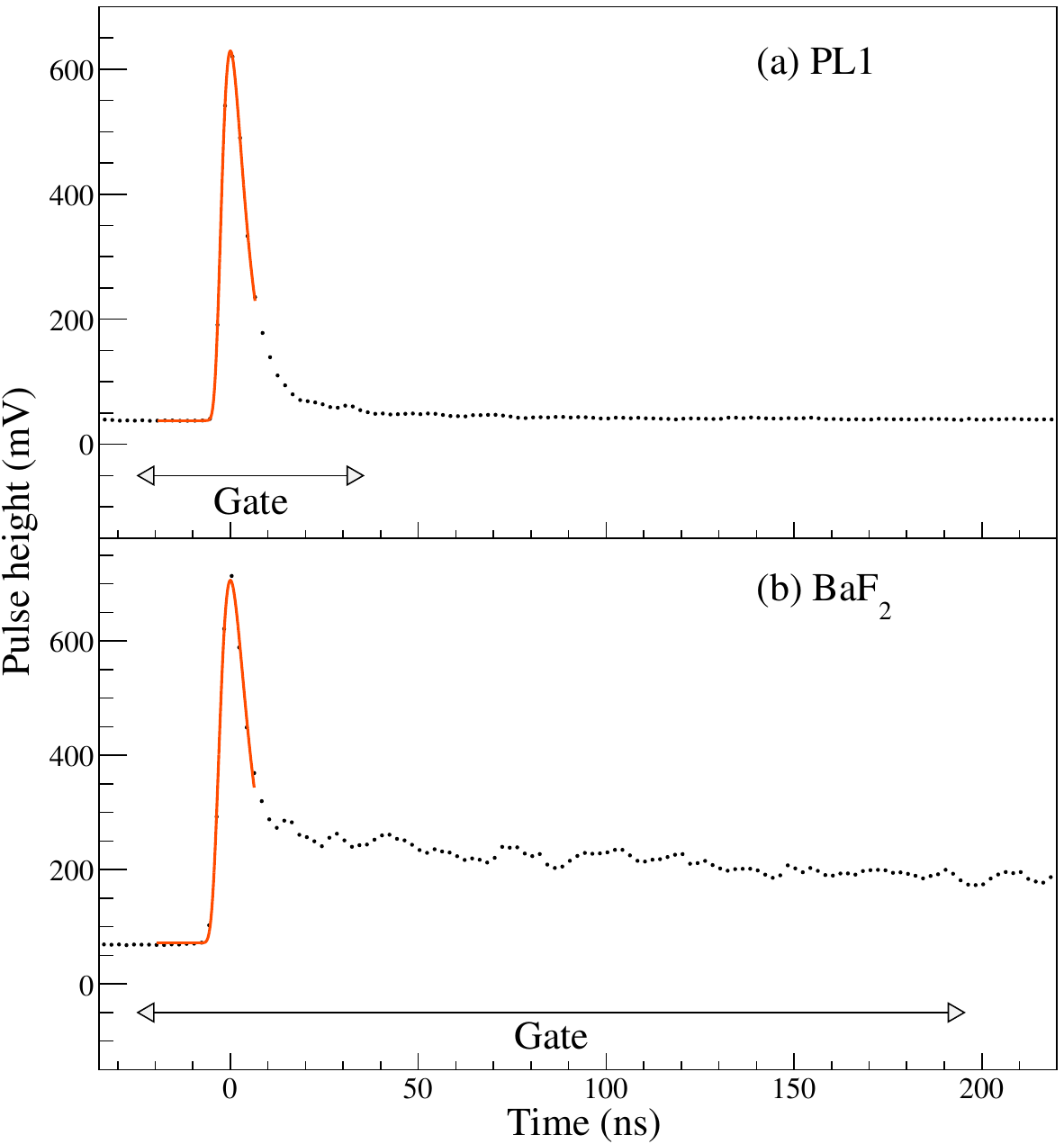}
  \caption{
  Typical waveforms of the output signal from (a) PL1 and (b) BaF$_2$ detectors.
  In both panels, the black dots represent the recorded waveform, and the red solid lines show the fitting function of Eq.~(\ref{eq:ana_commonfunction}). 
  The arrows indicate the charge integral gate. 
  The pulse shape of the BaF$_2$ detectors exhibits a long decay time remaining after the prompt pulse, which is the intrinsic feature of BaF$_2$ scintillator. 
  }
  \label{fig:ana_waveform}
\end{figure}

\begin{figure}
  \centering
  \includegraphics[width = 8.6cm]{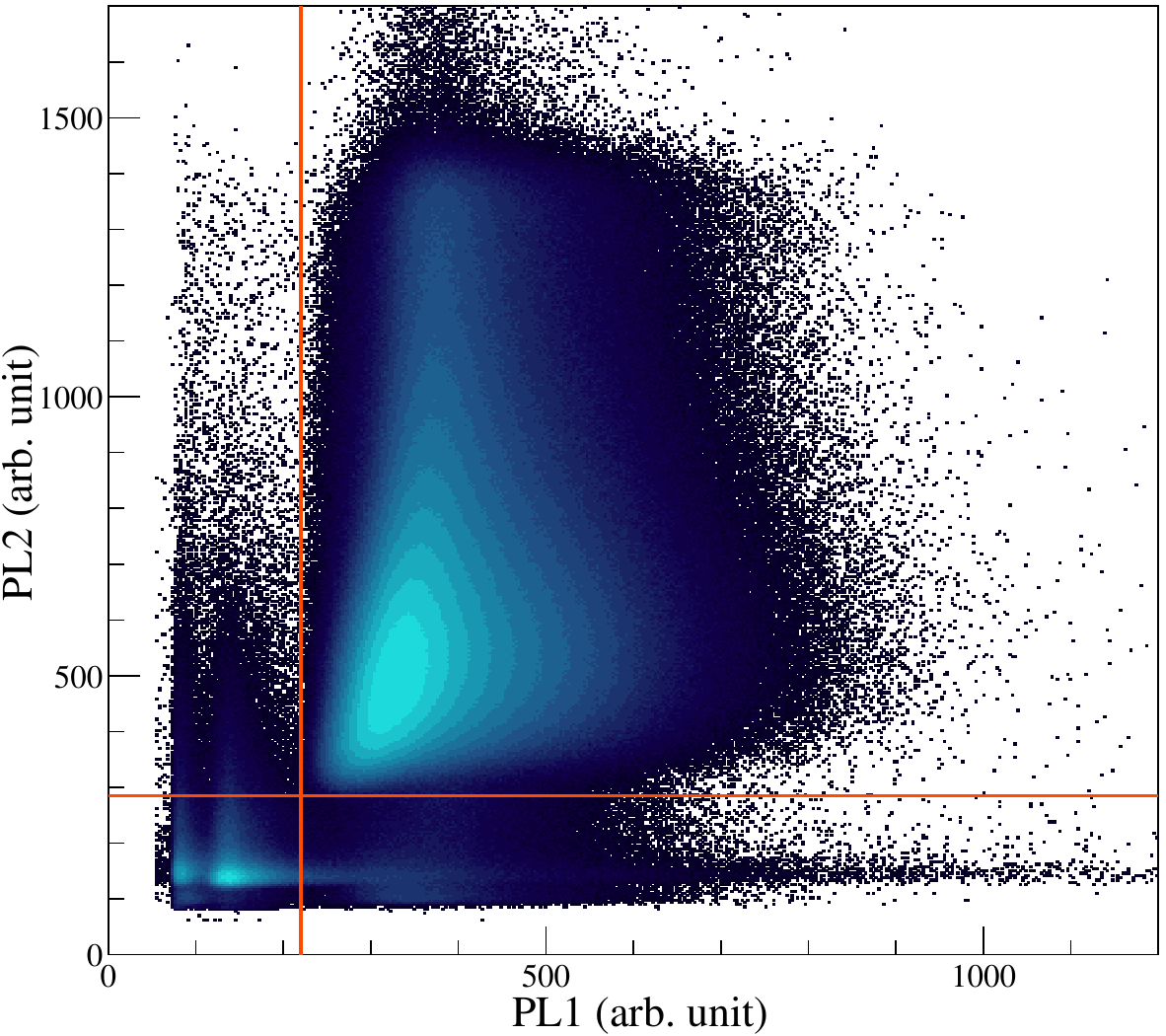}
  \caption{
  Two-dimensional energy-deposit histogram on PL1 and PL2.
  Muon events (upper right) are well separated from electron events (lower left).
  The red solid lines represent the threshold energies used for muon-electron discrimination.
  }
  \label{fig:ana_beampi}
\end{figure}
Beamline detectors were used to select the events of interest where the muon stopped in the palladium target.
Figure~\ref{fig:ana_waveform}(a) shows a typical signal waveform of PL1, with PL2 and PL3 having similar pulse shapes to PL1.
The energy deposit of the beam particle on these counters was deduced from the charge integration of the pulse within the integration gate, indicated with arrows in the figure.
The charge integration was calibrated to the energy deposit using the calculated energy loss for the muon and electron.
Figure~\ref{fig:ana_beampi} shows a two-dimensional histogram of the energy deposits in the upstream PL1 and PL2.
The upper right muon events are well separated from the lower left electron events.
The events above the thresholds, drawn with the red solid lines in the figure, were adopted as muon events.
Pion contamination was considered negligible due to the sufficient decay length of the MuSIC-M1 beamline.
Events in which a signal was detected on PL3, signifying that the beam particle had passed through the target, were excluded from the subsequent analysis.


\begin{figure}
  \centering
  \includegraphics[width = 8.6cm]{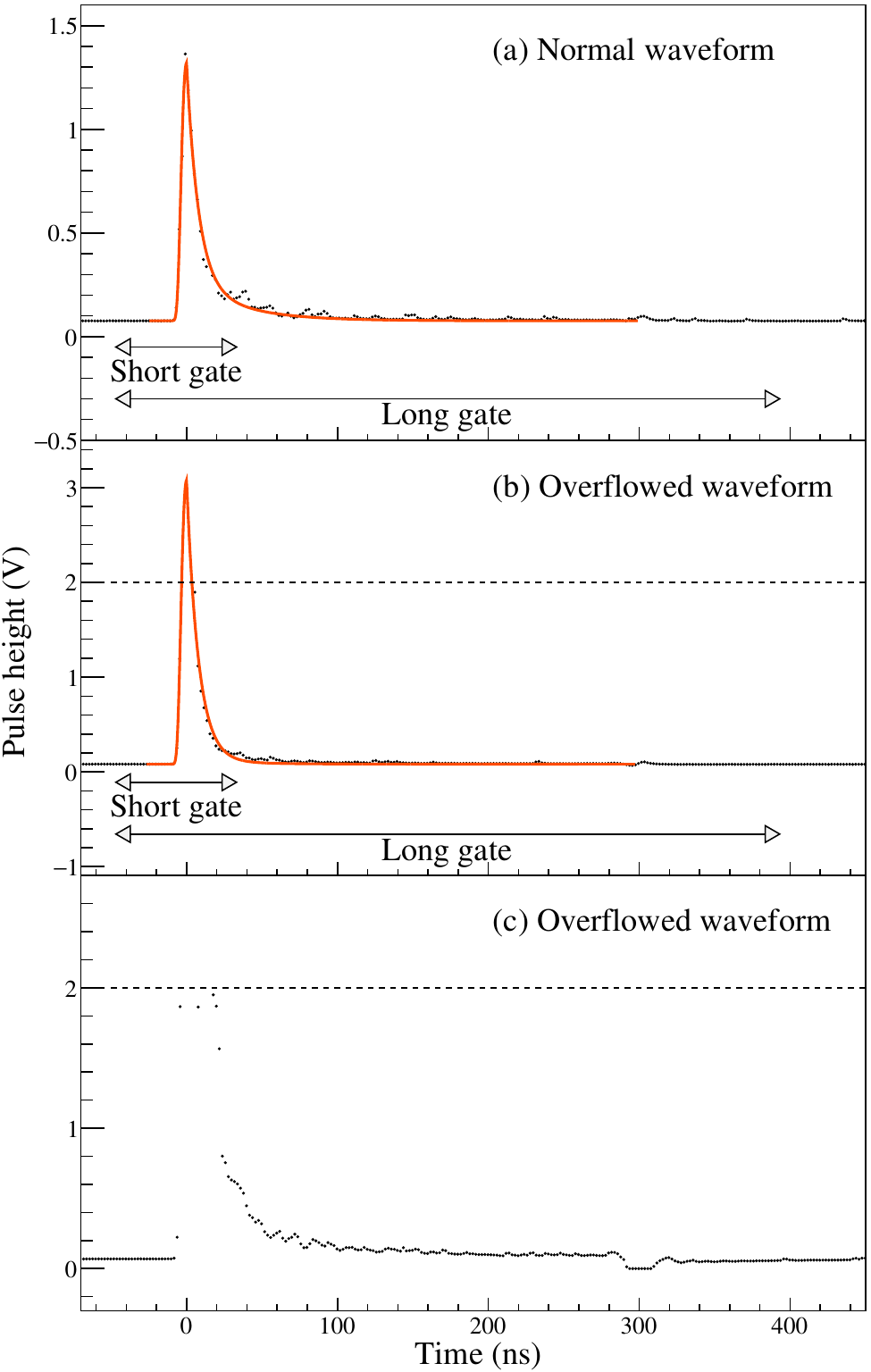}
  \caption{
  Typical waveform of the liquid scintillator with the fitting function (Eq.~\ref{eq:ana_liqfitfunction}) for different pulse heights:
  (a) Normal pulse height below the 2.0-V dynamic range of the digitizer,
  (b) Overflowed pulse that can be fitted by the function, and
  (c) Overflowed pulse that was excluded from the analysis.
  In all panels, the black dots represent the recorded waveform, and the red solid lines show the fitting function.
  The dotted lines in (b) and (c) represent the upper limit of the dynamic range of the digitizer.
  The arrows indicate the short and long gates for charge integration, which is used in the n-$\gamma$ discrimination analysis.
  }
  \label{fig:ana_liqwaveform}
\end{figure}
  
The timing of the beamline detectors and BaF$_2$ detectors was obtained by fitting the leading edge of the signal waveform, as shown in Fig.~\ref{fig:ana_waveform}.
The fitting function was:
\begin{align}
  &f_\mathrm{PL, BaF_2}(t) = H f_{0}(t) + B,\label{eq:ana_commonfunction} \\ 
  &f_{0}(t) = \exp{\left[ -\frac{1}{2} \left( \frac{t-t_0}{p} + \exp{\left( - \frac{t-t_0}{p} \right) } -1 \right)  \right] } ,
  \label{eq:ana_f0}
\end{align}   
where $H$ and $t_0$ are the maximum height and timing of the pulse, respectively, $p$ corresponds to the rise time, and $B$ is the baseline of the output signal.
This function reproduces well the waveform of both detectors within the fitting region.
The detection time for each detector was defined by $t_0$.
For PL2, the average of the two $t_0$ values read out from its two PMTs defines the detection time.
The detection time of PL2 was then used as the time origin in the following analysis.
For the liquid scintillators, the function containing both rising and decaying parts of the signal was used:
\begin{align}
 & f_{\rm liq} (t) = \begin{cases}
     H f_{0}(t) + B 
       \quad (t \leq t_0 ) \\
     H f_{1}(t) + B \quad (t>t_0 ) 
       \end{cases} 
       \label{eq:ana_liqfitfunction} \\
& f_{1}(t) = 
     \left[ (1 - r)  \exp{ \left(  - \frac{t-t_0}{\tau_1} \right) }  + r \exp{ \left(  - \frac{t-t_0}{\tau_2} \right) }   \right]  .
\end{align}
The rising part ($f_0(t)$) is the same as Eq.~(\ref{eq:ana_f0}), $H$ is the pulse height, and $t_0$ is the pulse timing.
The decaying part ($f_1(t)$) is a sum of two exponential functions with their decay constants of $\tau_1 = 3.9$\,ns and $\tau_2 = 17$\,ns.
These decay constants are the decay time of the fluorescent states of the liquid scintillator, taken from Ref.~\cite{Leo1994-jg}.
The parameter $r$ is the ratio of two components.
A typical signal waveform of the liquid scintillator with the fitting function of Eq.~(\ref{eq:ana_liqfitfunction}) is shown in Fig.~\ref{fig:ana_liqwaveform}(a).
The fitting function well reproduces the waveform for the whole fitting range.
The time offsets on $t_0$ for the BaF$_2$ detectors and liquid scintillators were deduced using the muonic X rays, emitted less than 10\,ps after the muon beam arrival.
The timing resolution of those detectors was estimated from the width of the prompt peak of the muonic X rays.
While the timing resolution has energy dependence, the typical timing resolutions were less than 2.5\,ns for the BaF$_2$ detector and less than 5\,ns for the liquid scintillator, respectively.


The energy deposit in the liquid scintillator was deduced from the numerical integration of the fitting function (Eq.~(\ref{eq:ana_liqfitfunction})) rather than direct waveform integration to ensure consistent treatment, particularly addressing the signal overflow described below.
The integration gate is shown in Fig.~\ref{fig:ana_liqwaveform} labeled as "Long gate".
The energy calibration of the liquid scintillators was performed in two steps.
First, energy calibration from the charge integral to the electron-equivalent (ee) energy was performed.
The characteristic L$_\alpha$ lines of the muonic palladium atom at 833 and 836\,keV were used as references.
Second, the electron-equivalent energy was converted into the proton-equivalent (pe) energy using the neutron energy obtained via the TOF method explained below, so that the proton-equivalent energy deposit reproduces the kinematics of the neutron.

\begin{figure}
   \includegraphics[width=86mm]{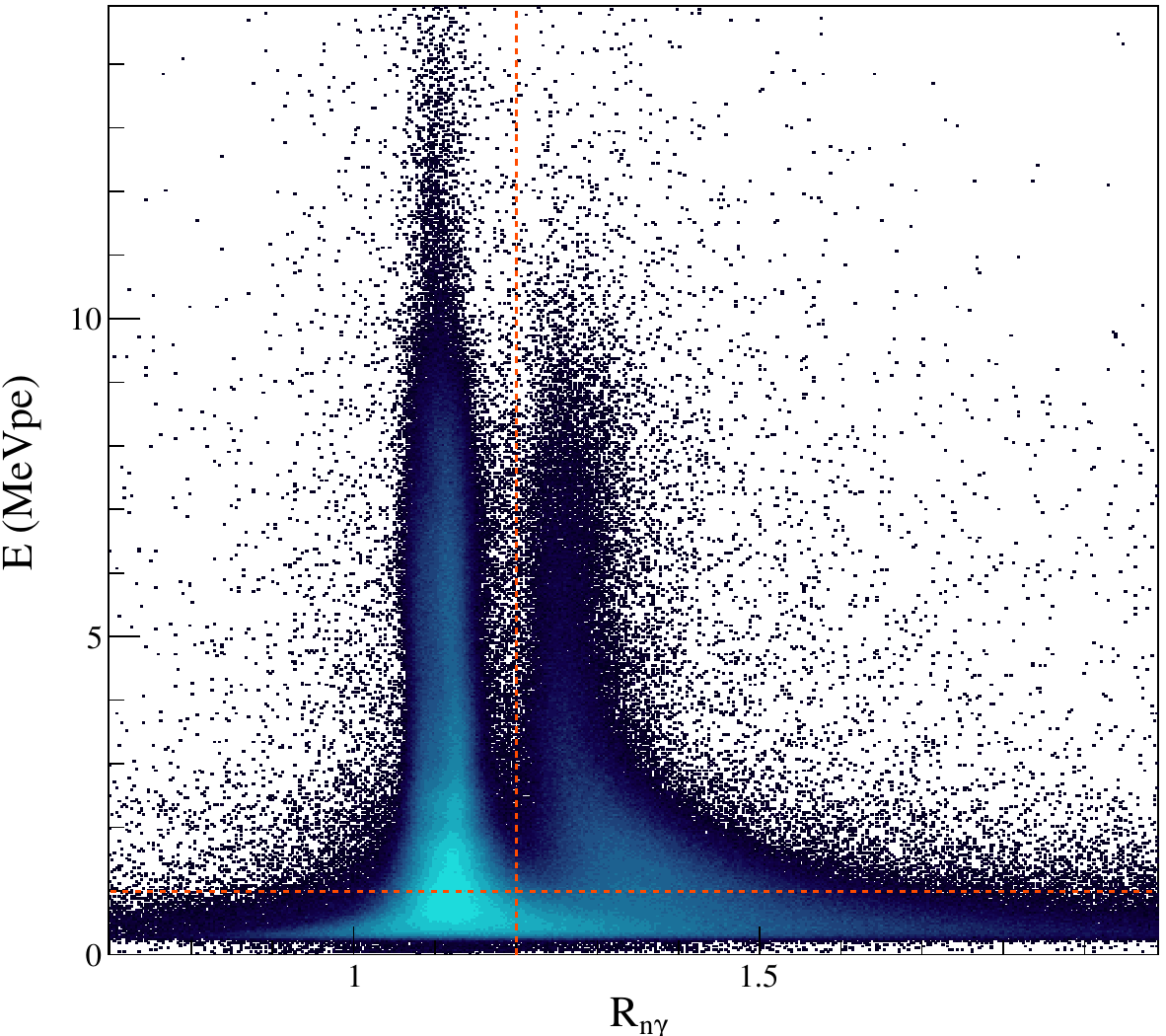}
   \caption{
   Pulse-shape discrimination between neutrons and $\gamma$ rays using the liquid scintillators.
   The x-axis represents the charge integral ratio defined by short and long integration gates as shown in Fig.~\ref{fig:ana_liqwaveform}, and the y-axis shows the proton-equivalent energy deposit in the scintillator.
   The red dotted lines represent the discrimination threshold (vertical) and energy-deposit threshold (horizontal).
   }
   \label{fig:ana_liqpi}  
\end{figure}

Some pulses from the liquid scintillators exceeded the 2.0-V dynamic range of the digitizer, referred to as a signal overflow.
Despite the overflows, the fitting procedure using Eq.~(\ref{eq:ana_liqfitfunction}) provided a good fit to the recorded region, as shown in Fig.~\ref{fig:ana_liqwaveform}(b), and properly extracted both the timing ($t_0$) and charge integral.
However, when the overflowed pulse height reached a certain level, a distortion of the recorded pulse was observed, as seen in Fig.~\ref{fig:ana_liqwaveform}(c).
These distorted pulses could not be recovered by the fitting procedure described above.
Such distortions may stem from a temporary malfunction of the waveform digitizer, likely caused by a charge buildup.
Consequently, these pulses were excluded from the analysis.
The exclusion of such pulses effectively sets an upper limit on the energy deposit measured by the liquid scintillator at 8\,MeV$_\mathrm{pe}$ for neutrons.

For the neutron-$\gamma$ discrimination using the pulse shape of the liquid scintillators, a ratio of the charge integral with short and long integration gates ($R_{n\gamma}$) was calculated.
The integration gates are illustrated with arrows in Figs.~\ref{fig:ana_liqwaveform}(a) and \ref{fig:ana_liqwaveform}(b).
Figure~\ref{fig:ana_liqpi} is a scatter plot of the pulse-shape discrimination.
Right neutron events were well separated from the left $\gamma$ rays above 1\,MeV$_\mathrm{pe}$.
The discrimination threshold, indicated by the red solid line in the figure, was used to select the neutron events.

The energy of the neutrons was calculated from their time-of-flight (TOF) and a neutron flight length of 40\,cm.
The TOF was deduced from the timing difference between the $\gamma$ rays measured by the BaF$_2$ detectors and neutrons measured by the liquid scintillators.
The prompt peak in the $\gamma$-ray spectra of the BaF$_2$ detectors, corresponding to the muonic X-ray, was excluded from this analysis.
The background estimated from the empty measurement was subtracted from the spectra.

\begin{figure}
   \includegraphics[width=86mm]{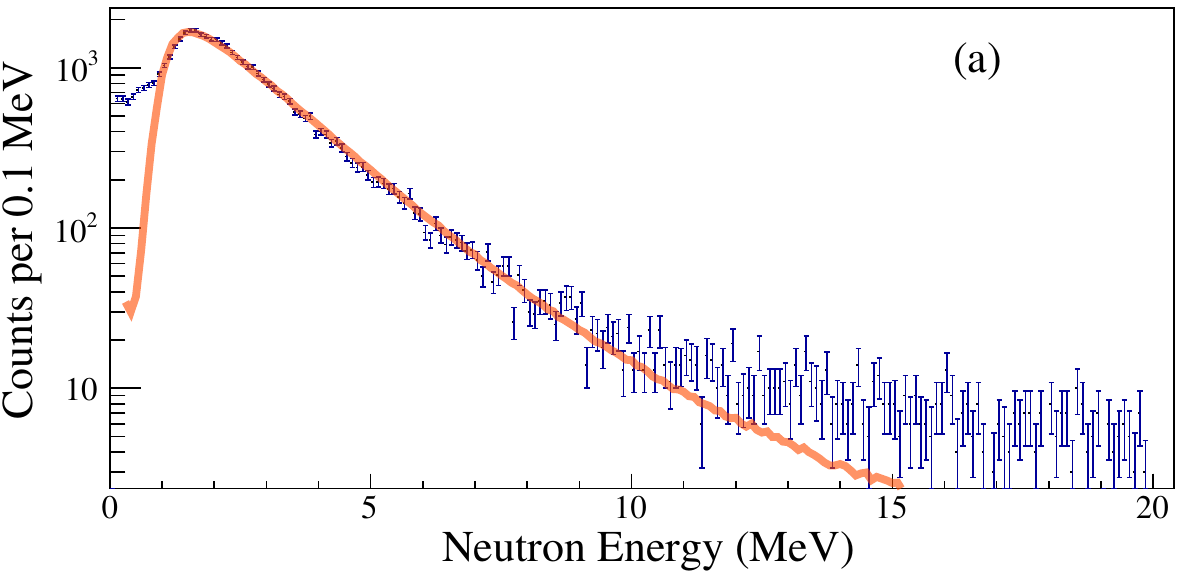}
   \includegraphics[width=86mm]{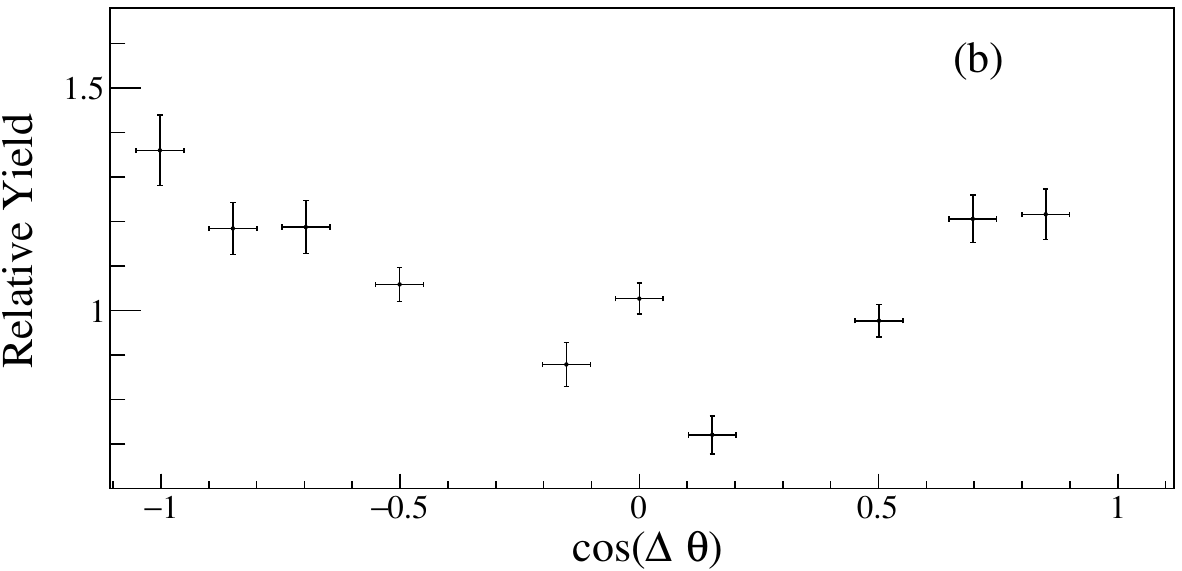}
   \caption{
    Neutron energy and opening angle spectra from the spontaneous fission of the $^{252}$Cf source.
    (a) Energy spectrum. 
    The red line represents a literature energy spectrum folded with the response function. The overall height of the literature curve was adjusted to the measured spectrum.
    (b) Relative opening angle distribution.
   }
   \label{fig:ana_Cfcalib}  
\end{figure}

The measured spectra resulted from the convolution of the intrinsic energy distribution with the detector's response function, which is determined by its energy resolution and detection efficiency.
The energy resolution was calculated by the timing resolution of the liquid scintillators and BaF$_2$ scintillators. 
For the liquid scintillators, the timing resolution was evaluated as a function of the energy deposit, while the averaged value against the $\gamma$-ray energy distribution was adopted for that of the BaF$_2$ scintillator.
The detection efficiency was estimated using a Monte Carlo simulation performed with the Particle and Heavy Ion Transport code System (PHITS, version 3.10)~\cite{Niita2006-xg, Sato2013-gq, Sato2024-xu, Sato2012-phits}.
The simulation used JENDL-4.0~\cite{Shibata2011-so} as the nuclear data library for neutron-induced reactions up to 20-MeV neutron kinetic energy, and EGS5~\cite{Hirayama2005-mo} for electron transport with electromagnetic interactions.
To validate the response function, a comparison was made with calibration data obtained from a $^{252}$Cf standard source.
This calibration measurement was performed over 23 hours in total with a source activity of approximately 2\,kBq, placed at the target position.
Figure~\ref{fig:ana_Cfcalib}(a) shows the neutron energy spectrum measured with the $^{252}$Cf source.
The neutron energy spectrum up to 10\,MeV produced by the spontaneous fission of $^{252}$Cf was well fitted with the literature curve \cite{Mannhart1989} convoluted by the response function.
The deviation above 10\,MeV in Fig.~\ref{fig:ana_Cfcalib}(a) is not identified. 
This might be attributed to detecting neutrons by the BaF$_2$ scintillators due to the high multiplicity of the fission.
The systematic uncertainty related to this will be discussed in the next section.


Events involving the simultaneous detection of two or more neutrons were utilized for the neutron-neutron angular correlation analysis.
For this analysis, the opening angle ($\Delta\theta$) between the two detected neutrons was deduced from the center positions of the neutron detectors relative to the target.
If three or more neutrons were coincidentally detected in one event, all possible combinations of two neutrons were considered in the analysis.
Monte Carlo simulations by PHITS were used to correct for the angular acceptance of the detector geometry considering the combination for every pair of detectors and the crosstalk effects from scattered neutrons.

Figure~\ref{fig:ana_Cfcalib}(b) shows the angular correlation of neutrons from $^{252}$Cf measured in the current setup. 
The pairs of neutrons produced with spontaneous fission of $^{252}$Cf have a characteristic angular correlation, and $\Delta \theta$ concentrates around forward (0$^{\circ}$) and backward (180$^{\circ}$)~\cite{Schuster2019}.
Excesses of around 0$^{\circ}$ and 180$^{\circ}$ were consistently observed in the current detector setup.
The ratios of the yield at 30$^{\circ}$ to that at 90$^{\circ}$ are 1.6 for this experiment and 1.67 for the previous study~\cite{Schuster2019}, respectively, which are in good agreement.

\section{Result and discussion}\label{sec_res}

\begin{figure*}
    \centering
    \includegraphics[width=18.0cm]{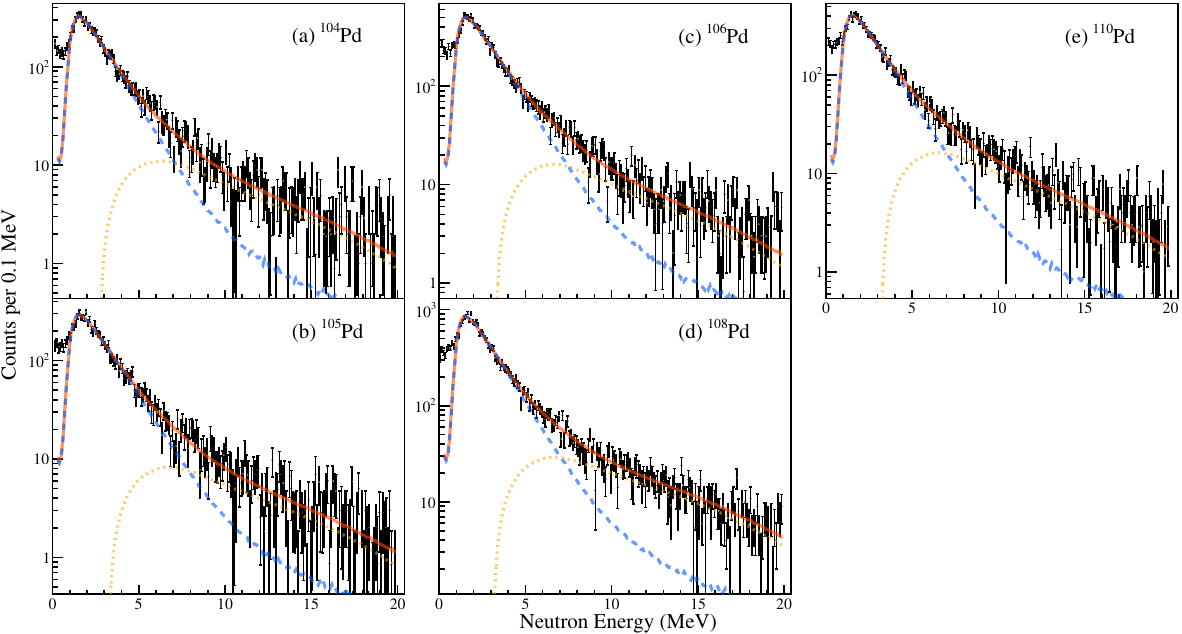}
    \caption{
        Neutron energy spectrum after nuclear muon capture on palladium isotopes with mass number $A=$ 104, 105, 106, 108, and 110.
        The black points represent the measured spectrum.
        The red solid and blue dashed lines are fitting with Eq.~(\ref{eq:res_fit}) and (\ref{eq:res_lowenergy}), respectively.
        The difference between the two fits, corresponding to the high-energy component, is displayed with the yellow dotted lines.
        The yields are not normalized.
    }
    \label{fig:nenergy}
\end{figure*}
Figure~\ref{fig:nenergy} shows the neutron energy spectra for nuclear muon capture on $^{104,105,106,108,110}$Pd measured in the present study. 
The energy spectra of the neutrons following muon capture have been measured for the nucleus around the $A=100$ region for the first time.

The neutron energy spectra after nuclear muon capture on natural thallium, lead, and bismuth targets have been measured by Schr\"{o}der et al.~\cite{Schroder1974-jb}.
The measured neutron energy spectrum was well reproduced by three characteristic spectral components, reflecting the direct, pre-equilibrium, and evaporation processes.
Hereafter, we will follow the analysis described there for a quantitative comparison.

The low-energy component of the neutron spectrum is dominated by the evaporation process. 
The energy spectrum resulting from the evaporation process is described~\cite{Le-Couteur1959-yb, Couteur1952-pw, Schroder1974-jb} as
\begin{equation}
 f_\mathrm{evap}(E) =  E^{5/11} \exp{\left( -\frac{E}{\Theta}\right) }
 \label{eq:res_evap},
\end{equation}
where $E$ and $\Theta$ are the neutron energy and a parameter corresponding to the temperature of the compound nucleus.
The dashed lines in Fig.~\ref{fig:nenergy} are the fitting results with the function of  
\begin{equation}
    \frac{\mathrm{d} N (E) }{\mathrm{d} E} = 
  N f_\mathrm{evap}(E) ,
  \label{eq:res_lowenergy}
\end{equation}
convoluted by the response function.
$N$ is a fitting parameter.
The fitting was performed between 1\,MeV and 4\,MeV as following the previous study~\cite{Schroder1974-jb}.
The parameter $\Theta$ obtained with Eq.~(\ref{eq:res_lowenergy}) is listed in Table~\ref{tab:res_energypara} for five palladium isotopes together with $\chi^2$ and the degrees of freedom (NDF) of the fitting.
In addition to the statistical uncertainty quoted in Table~\ref{tab:res_energypara}, the systematic uncertainty of $\Theta$ was estimated to be 0.06\,MeV by comparing the measured calibration spectrum of $^{252}$Cf with the literature and considering the dependence of the parameters in the analysis.

The high-energy neutrons are predominantly produced by the direct and pre-equilibrium processes.
While a model calculation was adopted in Ref.~\cite{Schroder1974-jb}, a phenomenological exponential function for the direct and pre-equilibrium components is assumed in the present study:
\begin{equation}
    f_\mathrm{he}(E) = \frac{1}{T (e^{-E_\mathrm{min}/T} - e^{-E_\mathrm{max}/T}) }  \exp{\left( - \frac{E}{T} \right) } \label{eq:res_he},
\end{equation}
where $T$ is an exponential scaling parameter, normalized between $E_\mathrm{min}$ and $E_\mathrm{max}$.
The solid line in Fig.~\ref{fig:nenergy} represents the fitting result for the function
\begin{equation}
 \frac{\mathrm{d} N (E) }{\mathrm{d} E} = 
  N \left[ f_\mathrm{evap}(E) +F f_\mathrm{he} (E)  \right].
 \label{eq:res_fit}
\end{equation}
convoluted with the response function.
$F$ is the fraction of high-energy component to the evaporation component, and $N$ is an overall factor for the fitting. 
The entire spectrum was fitted from 1 to 20\,MeV, with a condition of $E_\mathrm{min} = 4.5$\,MeV and $E_\mathrm{max} = 20$\,MeV.
The parameters $\Theta$, $T$, and $F$ obtained from the fitting by Eq.~(\ref{eq:res_fit}) are also listed in Table~\ref{tab:res_energypara}.

\begin{figure}
    \centering
    \includegraphics[width=8.6cm]{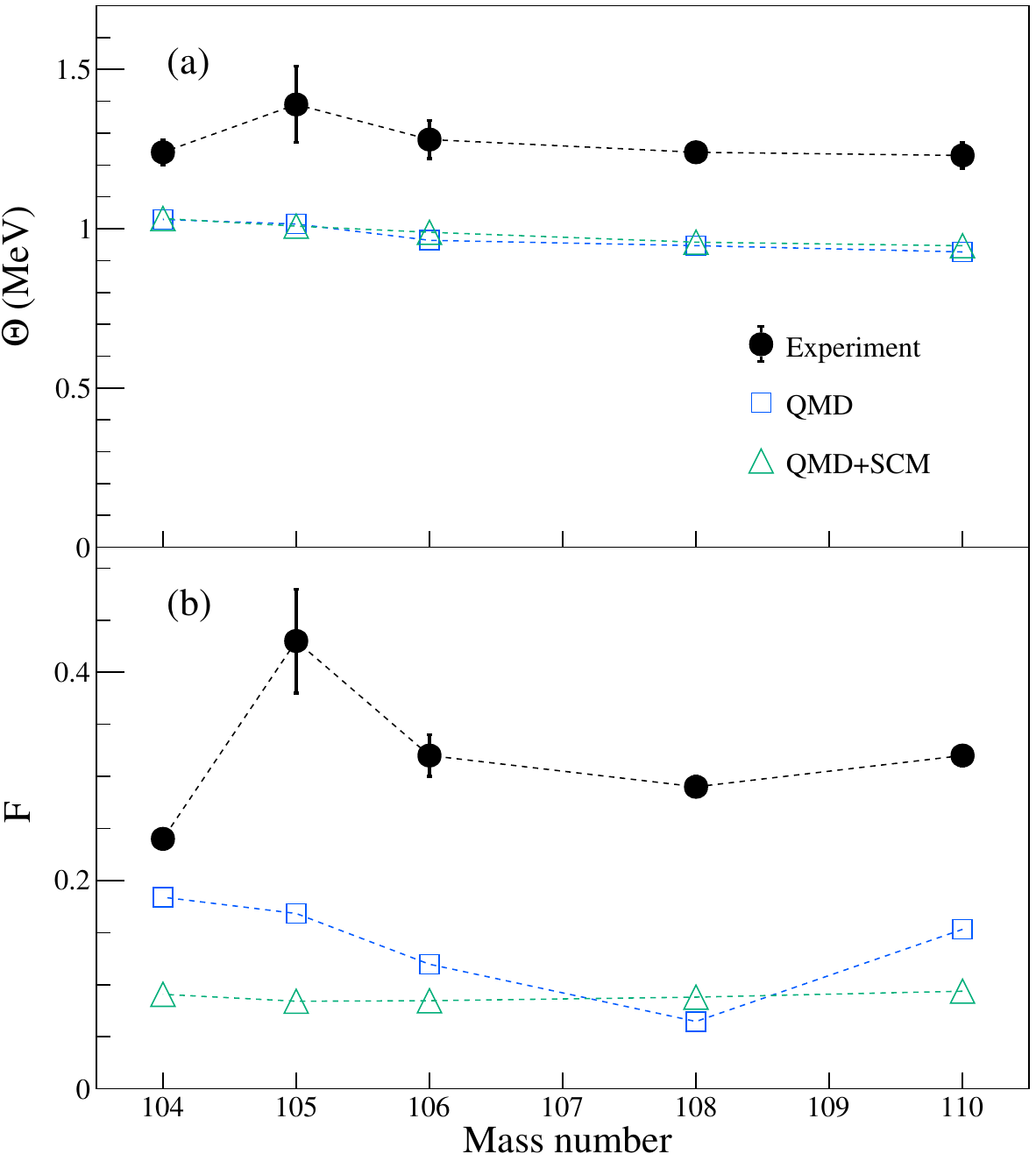}
    \caption{Isotope dependence of the parameters $\Theta$ and $F$ obtained by the fitting of the neutron spectrum. (a) $\Theta$ is from the fitting function Eq.~(\ref{eq:res_lowenergy}) and (b) $F$ is from Eq.~(\ref{eq:res_fit}).}
    \label{fig:res_isodep}
\end{figure}

The isotope dependence of the parameter $\Theta$ determined by the fitting with Eq.~(\ref{eq:res_evap}) is illustrated in Fig.~\ref{fig:res_isodep}(a).
The temperature of the compound nucleus produced by nuclear muon capture has no isotope dependence within the experimental uncertainty.

The average value of $\Theta$ for the palladium isotope, approximately 1.4\,MeV, is apparently larger than that for thallium (1.09\,MeV), lead (1.22\,MeV), and bismuth (1.06\,MeV)~\cite{Schroder1974-jb}.
The temperature $\Theta$ corresponds to the average excitation energy after nuclear muon capture $E^{*}$ and the density of the states $a$ as $\Theta \sim \sqrt{ ( E^{*}/a )}$, where $a = A/8$\,MeV$^{-1}$ for most nuclei.
The difference of $\Theta$ between the heavier nuclei around $A = 200 $ and the palladium nuclei around $A = 100$ could be explained by assuming the scaling against the mass difference through the simple $\Theta \sim \sqrt{ ( E^{*}/a )} =\sqrt{ ( 8E^{*}/A )} $ relationship.
This comparison suggests that the average excitation energy after nuclear muon capture is constant over the range from $A = 100$ to 200.
The estimated average excitation energy, $E^{*} \sim 25$\,MeV for $\Theta = 1.4$ and $A = 100$, is in good agreement with the typical estimated excitation energy of nuclear muon capture for medium-heavy nuclei~\cite{Measday2001-hi}.



\begin{table}
    \centering
    \caption{
        Summary of fitting parameters $\Theta$, $T$, $F$, and $\chi^2$/NDF for neutron energy spectra.
        Numbers in parentheses indicate the statistical uncertainties.
        The units for $T$ and $\Theta$ are MeV.
        The systematic uncertainty is $\pm$0.06~MeV for $\Theta$.
    }
    \label{tab:res_energypara}
    \begin{tabular}{cccccccc} 
    \hline \hline
        & \multicolumn{2}{c}{Eq.~(\ref{eq:res_lowenergy})} & & \multicolumn{4}{c}{Eq.~(\ref{eq:res_fit}) }   \\
    \cline{2-3} \cline{5-8}  
        A   & $\Theta$ & $\chi^2$/NDF & & $\Theta$ & $T$    & $F$     & $\chi^2$/NDF \\
    \hline
        104 & 1.34(3)  & 29.4/28      & & 1.25(4)  & 15(5)  & 0.25(1) & 186/185      \\
        105 & 1.46(4)  & 35.3/28      & & 1.35(5)  & 17(7)  & 0.26(2) & 175/185      \\
        106 & 1.42(3)  & 43.5/28      & & 1.29(3)  & 15(4)  & 0.27(1) & 230/185      \\  
        108 & 1.35(2)  & 72.8/28      & & 1.24(2)  & 35(14) & 0.30(1) & 226/185      \\
        110 & 1.41(3)  & 27.7/28      & & 1.25(4)  & 13(3)  & 0.32(1) & 195/185      \\
    \hline \hline
    \end{tabular}
\end{table}
 


Most of the previous measurements of neutron energy after nuclear muon capture focused on the high-energy region, typically exceeding 20\,MeV~\cite{Sundelin1968-bg, Plett1971-dm, Sundelin1973-xg}.
On the other hand, measurement of the neutrons around 10\,MeV and derivation of the fraction $F$ are crucial for discussing the transition from the direct and pre-equilibrium processes to the evaporation process.
To date, experimental results focusing on this region have only been reported by Ref.~\cite{Schroder1974-jb} and the present study.
In addition, the present study presents the first experimental results that addresses the isotope dependence of $F$.
As shown in Fig.~\ref{fig:res_isodep}(b), there is no significant isotope dependence of $F$ beyond the statistical uncertainty for even isotopes; however, $^{105}$Pd exhibits a slightly larger value than the other even isotopes.
Although the present result may imply an odd-even effect in $F$, we refrain from drawing a definitive conclusion given only one experimental point for an odd isotope.

The experimental results were compared with a muon interaction model implemented in PHITS~\cite{Abe2017-xw}.
The model consists of three steps.
First, the energy of the neutrons produced by nuclear muon capture is sampled based on the excitation energy distribution model proposed by Singer~\cite{Singer1962-ww}, in which the momentum distribution of the proton in the nuclear medium was estimated using the Amado model~\cite{Amado1976-cy}.
The semiclassical transportation and scattering process of the neutrons by other nucleons was simulated by the quantum molecular dynamics (QMD)~\cite{Niita1995-jqmd, Ogawa2015-gu}.
The scattered nucleons that exit the nucleus during the QMD calculation correspond to the direct and pre-equilibrium process.
After a certain cutoff time, the generalized evaporation model (GEM)~\cite{Furihata2000-nl} was used to calculate the evaporation process.
The PHITS calculations were performed using two models: the default QMD model (QMD) and the one modified by incorporating the Surface Coalescence Model (QMD+SCM)~\cite{Watanabe2007-mm}.
The SCM was introduced in PHITS to reproduce the emission probability for light complex particles at high energies for nuclear muon capture of Si isotopes~\cite{Manabe2023-zx}.
The energy spectrum of neutrons obtained by the PHITS calculation was fitted by the functions of Eqs.~(\ref{eq:res_fit}) and (\ref{eq:res_lowenergy}) over the same fitting regions as the experiments.
The parameters $\Theta$ and $F$ obtained by this procedure are illustrated in Fig.~\ref{fig:res_isodep}.

The PHITS calculation underestimated the parameter $\Theta$ in the evaporation function of Eq. (\ref{eq:res_lowenergy}) by approximately 1.3 times, while two calculation models used in the present study resulted almost the same trend in $\Theta$. 
The underestimation of the temperature $\Theta$ corresponds to the underestimation of the energy in the GEM sequence.
In contrast, our previous study of the production branching ratio following nuclear muon capture indicates the overestimation of the neutron multiplicity by the PHITS calculation and suggested overestimation of the excitation energy by the Singer model~\cite{Niikura2024-bz}.
This comparison implies that the simple parametric optimization of the excitation energy cannot reproduce both sets of experimental data.

While two models in PHITS exhibit slightly different values of the fraction $F$, all calculated values were underestimated by a factor of approximately two compared to the experimental ones. 
This underestimation indicates that the probability of high-energy neutron emission by the QMD, which calculates the direct and pre-equilibrium processes, is insufficient to reproduce the experimental spectra.
A previous study on the production branching ratio~\cite{Niikura2024-bz} revealed that PHITS calculations underestimated the single neutron emission probability. 
Since neutrons emitted from direct and pre-equilibrium processes typically possess high energies and mostly result in the emission of only a single neutron, this underestimation of the single neutron emission probability was interpreted as reflecting an underestimation of the direct and pre-equilibrium components.
The present neutron energy measurement strongly supports this interpretation.
As $F$ directly reflects the ratio of direct and pre-equilibrium to evaporation processes, the neutron energy measurement around 10 MeV offers a direct insight into these reaction dynamics, thereby highlighting its crucial importance for understanding nuclear muon capture.

\begin{figure*}
    \centering
    \includegraphics[width=18.0cm]{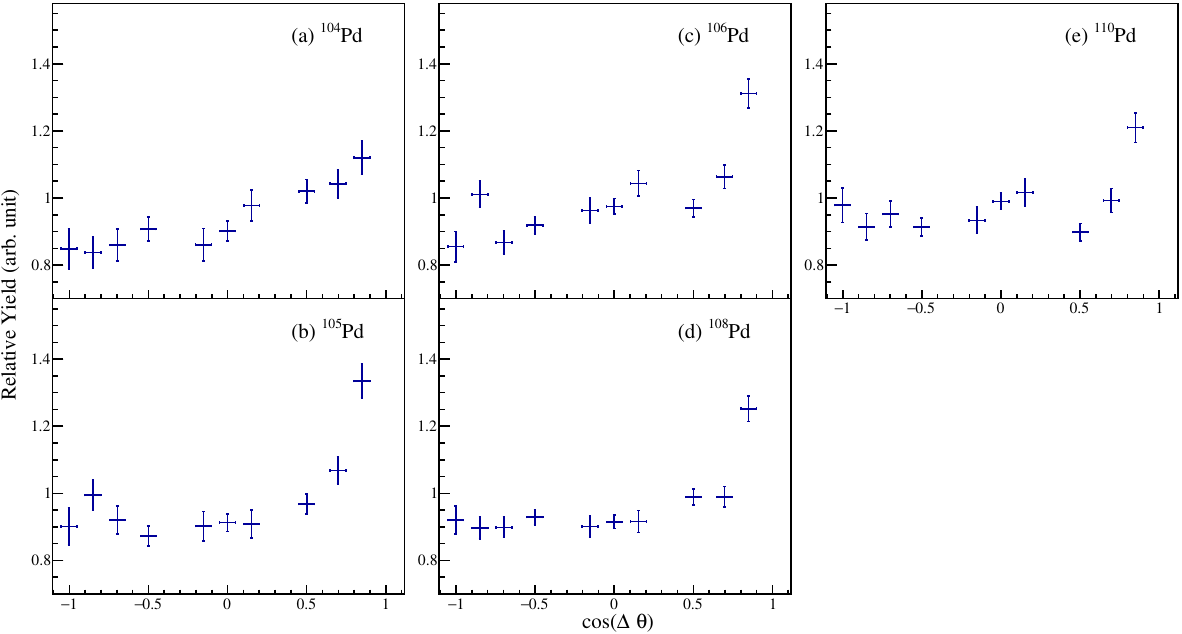}
    \caption{Opening-angle correlation of two neutrons emitted following nuclear muon capture on palladium isotopes. The detector response, such as efficiency, acceptance, and crosstalk probability, is corrected so that random neutron produce the relative yield of 1. }
    \label{fig:angle}
\end{figure*}

Figure~\ref{fig:angle} shows the opening angle ($\Delta \theta$) distributions of two neutrons after the correction of the angular acceptance of the detector.
The uncertainty in the figure is only a statistical one.
While there are the small differences among the isotopes, clear excesses at the small opening angle ( $ < 60^{\circ}$) can be found for all isotopes.
The neutrons detected in the current experiment are dominated by the low energy neutrons as shown as Fig.~\ref{fig:nenergy}, and thus the angular correlation is considered to be for the low energy neutrons.
The experimental biases, such as crosstalk and efficiency, were carefully considered, and is smaller than the statistical uncertainty comparing the $^{252}$Cf measurements as shown in Fig.~\ref{fig:ana_Cfcalib}(b).


There is no previous study for the neutron angular correlation after nuclear muon capture for the opening angle smaller than $90^{\circ}$.
The possible comparison can be found in  Ref.~\cite{Kozlowski1985-lb}.
The neutron-neutron correlation at the large opening angle (180$^{\circ}$) is reported there for the neutrons with a larger energy deposit than 10~MeV following nuclear muon capture on $^{40}$Ca.
The correlation at the large opening angle is due to the virtual pion production during nuclear muon capture.
This correlation at the large opening angle thus corresponds to the direct component of the neutrons.
On the other hand, the correlation at the small opening angle observed in the present measurement is for the low-energy evaporation neutrons.
The calculation by QMD+GEM was also performed, resulted an almost flat distribution.
Although the nucleon-nucleon correlation in the ground state, dynamics in the pre-equilibrium stage, and quantum interference~\cite{Hanbury-Brown1956-dg} could explain the excess at the small $\Delta \theta$, the reasons for the neutron correlation obtained in the present study are currently not clear.


\section{Conclusion}\label{sec_con}

In the present study, we measured the neutron energy spectra following nuclear muon capture on isotopically enriched palladium isotopes ($A=$ 104, 105, 106, 108, and 110). The experiment was carryed out on the MuSIC-M1 beamline at RCNP, Osaka University. An array of liquid scintillators and BaF$_{2}$ detectors was utilized for neutron and $\gamma$-ray detection, respectively, and neutron energy was deduced with the time-of-flight method using the start time defined by the $\gamma$-ray detection.

From the analysis of the measured neutron energy spectra, we derived the temperature parameter $\Theta$ of the compound nucleus produced by nuclear muon capture and the fraction $F$ of the direct and pre-equilibrium processes relative to the evaporation process. 
The obtained $\Theta$ values were found to be consistent with heavier nuclei by introducing the mass scaling rule, suggesting a relatively constant average excitation energy induced by muon capture across the mass range. 
Regarding the fraction $F$, our results indicate no significant isotope dependence for even palladium isotopes. 
An excess at small opening angles was also observed in the low-energy neutron-neutron angular correlation.

The experimental results were compared with model calculations performed using the PHITS code. 
The comparison provides crucial input for refining theoretical models, particularly in accurately describing the deexcitation dynamics following nuclear muon capture.
The measurement of the neutron energy is important as they provide quantitative insights about the temperature of the compound nucleus and the contributions of direct, pre-equilibrium, and evaporation processes. 

\section*{Acknowledgements}

The authors would like to acknowledge the accelerator team at RCNP, Osaka University, for providing the stable proton primary beam during the experiment.
We thank Prof.~Igashira and Prof.~Katabuchi for providing the isotope-enriched palladium targets.
This work is funded by the ImPACT Program of the Council for Science, Technology, and Innovation (Cabinet Office, Government of Japan) and partially supported by JSPS KAKENHI Grant Number JP18J10554.
T.Y.S. acknowledges support obtained from the ALPS Program at the University of Tokyo.

\bibliography{paperpile2}

\end{document}